\documentclass[a4paper,12pt]{article}
\pdfoutput=1
\usepackage{amssymb}
\usepackage{amsmath}
\usepackage{mathtools}
\usepackage{slashed}
\usepackage{color}
\usepackage{indentfirst}
\usepackage{graphicx}
\usepackage{bbm}
\usepackage{csquotes} %fix backwards quotes
\usepackage{caption}
\usepackage{cite}
\renewcommand{\thefootnote}{\fnsymbol{footnote}}

\begin{document}

\title{
\begin{flushright}
\begin{minipage}{0.2\linewidth}
\normalsize
EPHOU-20-0002\\
KEK-TH-2191 \\*[50pt]
\end{minipage}
\end{flushright}
{\Large \bf 
Common origin of the strong CP and CKM phases\\ in string compactifications
\\*[20pt]}}

\author{Tatsuo Kobayashi$^{a}$\footnote{
E-mail address: kobayashi@particle.sci.hokudai.ac.jp
}
\ and\
Hajime~Otsuka$^{b}$\footnote{
E-mail address: hotsuka@post.kek.jp
}\\*[20pt]
$^a${\it \normalsize 
Department of Physics, Hokkaido University, Sapporo 060-0810, Japan} \\
$^b${\it \normalsize 
KEK Theory Center, Institute of Particle and Nuclear Studies, KEK,}\\
{\it \normalsize 1-1 Oho, Tsukuba, Ibaraki 305-0801, Japan}}
\maketitle

\date{
\centerline{\small \bf Abstract}
\begin{minipage}{0.9\linewidth}
\medskip 
\medskip 
\small
We explore the scenario where both the strong CP and Cabbibo-Kobayashi-Maskawa (CKM) phases are determined by the same 
axion field. Such a scenario is naturally realized in string compactifications. We find that there exists parameter region to 
realize the tiny strong CP phase and observed CKM phase in magnetized 
D-brane models. 
\end{minipage}
}

\renewcommand{\thefootnote}{\arabic{footnote}}
\setcounter{footnote}{0}
%\vspace{35pt}
\thispagestyle{empty}
\clearpage
\addtocounter{page}{-1}
%\newpage
%\setcounter{tocdepth}{3}

\tableofcontents

%%%%%%%%%%%%%%%%%%%%%%%%%%%%%%%%%%%%%%%%%%%%%%%%%%%%%%%%%%%%
%%%%%%%%%%%%%%%%%%%%%%%%%%%%%%%%%%%%%%%%%%%%%%%%%%%%%%%%%%%%
\section{Introduction}\label{sec:introduction}
%%%%%%%%%%%%%%%%%%%%%%%%%%%%%%%%%%%%%%%%%%%%%%%%%%%%%%%%%%%%
%%%%%%%%%%%%%%%%%%%%%%%%%%%%%%%%%%%%%%%%%%%%%%%%%%%%%%%%%%%%

In the standard model, the strong CP phase $\theta_{\rm eff}=\theta_{\rm QCD} +{\rm Arg}\{{\rm Det}(y_uy_d)\}$ is severely constrained by experiments to be 
smaller than $10^{-10}$~\cite{Harris:1999jx,Baker:2006ts,Afach:2015sja}, 
but the CP violating CKM phase is 
of ${\cal O}(1)$. This strong CP problem indicates a 
new mechanism to solve not accommodated in the standard model. 
The most promising scenario solving the strong CP problem is 
to introduce the QCD axion which is a pseudo Nambu-Goldstone 
boson associated with the Peccei-Quinn (PQ) symmetry~\cite{Peccei:1977hh}. 
The strong CP phase $\theta_{\rm eff}$ can be determined by 
a dynamics of the axion field.

When we consider the string theory as well as higher-dimensional theory as 
an ultraviolet completion of the standard model, promoting the CP phase to the axion is naturally realized. 
Indeed, in the string theory, all the couplings are functions of moduli fields whose vacuum expectation values determine the size of the couplings in the standard model. 
In particular, both the QCD phase $\theta_{\rm QCD}$ as well as ${\rm Arg}\{{\rm Det}(y_uy_d)\}$ are functions of axion fields in general. 
It is then interesting to ask whether the QCD and CKM phases have a common origin in string compactifications and at the same time the tiny strong CP phase $\theta_{\rm eff}\ll 1$ is compatible with the ${\cal O}(1)$ CKM phase. 

In the low-energy effective action of string theory, 
the gauge kinetic function determining the QCD phase is linearly 
dependent of the moduli fields, whereas the axion dependence of the Yukawa couplings is model dependent in general. 
If the Yukawa couplings have the Froggatt-Nielsen (FN) type structure~\cite{Froggatt:1978nt} 
to realize the hierarchical structure of the fermion masses, $y_u\propto e^{2\pi i (k_{Q_i} +k_{U_j})a}$ $y_d\propto e^{2\pi i (k_{Q_i} +k_{D_j})a}$ with $k_{Q_i},k_{U_j},k_{D_j}$ being 
the FN charges of the quarks and $a$ the axion, the CKM phase is induced by the nonvanishing field value of the axion $a$. 
Furthermore, ${\rm Arg}\{{\rm Det}(y_uy_d)\}$ is also linearly 
dependent of the axion, namely ${\rm Arg}\{{\rm Det}(y_uy_d)\}\propto a$. 
It indicates that when the axion has a common origin of the QCD and CKM phases, $\theta_{\rm eff}=0$ requires $\langle a\rangle=0$. 
Then, the CKM phase cannot be generated. 
In this way, a certain non-trivial axion dependence to the Yukawa couplings is required to realize $\theta_{\rm eff}\ll 1$ compatible with the ${\cal O}(1)$ CKM phase. 
In this paper, we resolve this issue in a specific string compactification and explore the parameter region in the moduli space of the axion field to realize such a scenario. 
It gives a new insight in the strong CP problem from the view point of string theory.

This paper is organized as follows. 
In section 2, we show the origin of the QCD and CKM phases in Type IIB string 
on toroidal orientifold with D3/D7-branes. 
In particular, in Type IIB flux vacua, it is possible to have a relation between 
the QCD and CKM phases. 
In section 3, we explicitly evaluate both phases in a specific three-generation 
model realized in magnetized D7-branes and explore the parameter region leading 
to the observed CKM phase consistent with the tiny strong CP phase. 
It turns out that the axion controlling the magnitude of both phases provides 
semi-realistic observed values, namely mass ratios for the quarks, the elements of the CKM matrix and 
the Jarlskog invariant, thanks to the non-trivial axion-dependent Yukawa couplings. 
Section 4 is devoted to the conclusions.

%%%%%%%%%%%%%%%%%%%%%%%%%%%%%%%%%%%%%%%%%%%%%%%%%%%%%%%%%%%%%%%%%%%%%%%%%%%%%%%%
\section{The Model}
%%%%%%%%%%%%%%%%%%%%%%%%%%%%%%%%%%%%%%%%%%%%%%%%%%%%%%%%%%%%%%%%%%%%%%%%%%%%%%%%

In this section, we show how to relate the QCD phase with 
the CKM phase in the effective action of superstring theory 
with an emphasis on Type IIB string theory on toroidal orientifold 
$\Pi_{i=1}^3(T^2)_i/(\mathbb{Z}_2\times \mathbb{Z}_2)$ with D3/D7-branes. 

%%%%%%%%%%%%%%%%%%%%%%%%%%%%%%%%%%%%%%%%%%%%%%%%%%%%%%%%%%%%%%%%%%%%%%%%%%%%%%%%
\subsection{Origin of the QCD and CKM phases}
%%%%%%%%%%%%%%%%%%%%%%%%%%%%%%%%%%%%%%%%%%%%%%%%%%%%%%%%%%%%%%%%%%%%%%%%%%%%%%%%

Let us first consider the origin of the QCD phase on gauge fields living on the magnetized D$7_a$-brane wrapping the 4-cycle $(T^2)_j\times (T^2)_k$ with $j\neq  k$, where the  U(1) magnetic fluxes $F_a$ are introduced as
\begin{align}
    \frac{m_a^j}{l_s^2}\int_{(T^2)_j}F_a^j =n_a^j,
\end{align}
where $l_s=2\pi\sqrt{\alpha^\prime}$ is the string length. 
Here, $m_a^j$ and $n_a^j$ are the wrapping number of D$7_a$-brane and the quantized flux, respectively. 
The gauge kinetic function on the magnetized D$7_a$-brane is given by~\cite{Lust:2004cx, Lust:2004fi, Lust:2004dn}
\begin{align}
    f_{{\rm D}7_a} =|m_a^km_a^l|\left(T^i -\frac{n_a^k}{m_a^k}\frac{n_a^l}{m_a^l}\tau\right),
    \qquad (i\neq j\neq k)
\label{eq:fD7}
\end{align}
from which the CP phase is determined by two axions, 
originating from the Ramond-Ramond 4-form ${\rm Re}(T^i)= \int_{(T^2)_j\times (T^2)_k} C_4$ 
and Ramond-Ramond 0-form ${\rm Re}(\tau)= C_0$, respectively. 
The imaginary part of the K\"ahler moduli $T_i$ now denotes the volume of four-cycle  wrapped by the D7-brane, following \cite{Lust:2004dn}.

Next, we discuss the origin of the CKM phase in Yukawa couplings of matter fields living on magnetized D7-branes. 
Let us consider $U(N)$ magnetic flux on $N$ stacks of D7-branes 
such that $U(N)$ gauge symmetry on D7-branes is broken to $U(N_a)\times U(N_b)\times U(N_c)$ with $N=N_a+N_b+N_c$. 
Thanks to the magnetic fluxes, bifundamental zero-modes for $(N_\alpha, \bar{N}_\beta)$, $\alpha,\beta =a,b,c$ have the 
net number of index labeled by $p=0,1,\cdots, |I_{\alpha \beta}^j|-1$ 
with $I_{\alpha \beta}^j=n_\alpha^j/m_\alpha^j-n_\beta^j/m_\beta^j$ on 
each 2-torus $(T^2)_j$ wrapped by D7-branes 
and these degenerate chiral zero-modes can be identified with the quarks and/or leptons. 
From the analysis in the low-energy effective action of magnetized D7-branes, 
Yukawa couplings of such chiral zero-modes are 
found by calculating the overlap integral of zero-mode wavefunctions. 
On each 2-torus $(T^2)_j$ inside the 4-cycle wrapped by magnetized D7-branes, holomorphic Yukawa couplings are provided by~\cite{Cremades:2004wa}
\begin{align}
Y_{pqs} 
&= 
%\left|\frac{I_{\alpha \beta}^jI_{\gamma \alpha}^j}{I_{\beta \gamma}^j}\right|^{1/4}
\vartheta
\begin{bmatrix}
-\frac{1}{I_{ab}^j}\left(\frac{q}{I_{ca}^j}+\frac{s}{I_{bc}^j}\right) \\ 0
\end{bmatrix}
\left(0,\tau_j \left|I_{ab}^jI_{bc}^jI_{ca}^j\right|\right)
,
\label{eq:Yukawa}
\end{align}
up to the normalization factor, 
where $q=0,1,\cdots, |I_{ca}^j|-1$, $s=0,1,\cdots, |I_{bc}^j|-1$\footnote{Here, we employ $p=s-q$ mod $I_{ab}$. 
(For more details, see, Ref.~\cite{Cremades:2004wa}.)} and 
$\vartheta$ denotes the Jacobi theta function as 
a function of the complex structure modulus $\tau_j$
\begin{align}
\vartheta
\begin{bmatrix}
c\\
0
\end{bmatrix}
(0, \tau_j)
\equiv 
\sum_{l\in \mathbb{Z}}
e^{\pi i (c+l)^2\tau_j}
.
\end{align}
Note that a higher-dimensional gauge coupling as well as the K\"ahler metric of the 
matter field are involved in the physical Yukawa couplings, but they 
are real. 
In this way, the CKM phase is determined by the real part of complex structure moduli and 
the dependence of the axion field is not a Froggatt-Nielsen type as explicitly analyzed in a concrete model in section~\ref{sec:3}. 
The above calculation can be extended to the $T^2/\mathbb{Z}_N$ 
orbifold, where the Yukawa couplings are provided by the linear combination of Yukawa couplings. (For more details, see Ref.~\cite{Abe:2008fi}.)
In the next section, we propose the mechanism to correlate the QCD phase to the CKM phase.

%%%%%%%%%%%%%%%%%%%%%%%%%%%%%%%%%%%%%%%%%%%%%%%%%%%%%%%%%%%%%%%%%%%%%%%%%%%%%%%%
\subsection{Relation between the QCD and CKM phases}
%%%%%%%%%%%%%%%%%%%%%%%%%%%%%%%%%%%%%%%%%%%%%%%%%%%%%%%%%%%%%%%%%%%%%%%%%%%%%%%%

From now on, we show one of the mechanisms to relate the QCD phase with the CKM phase 
in Type IIB flux compactification on $T^6/(\mathbb{Z}_2\times \mathbb{Z}_2)$ with 
hodge numbers $(h_{1,1}, h_{2,1})=(3, 51)$. 
The closed string moduli in this setup are the axio-dilaton $\tau$, three K\"ahler moduli $T_i$ and 
untwisted complex structure moduli $\tau_j$ with $i,j=1,2,3$. 
The K\"ahler potential of moduli fields are described by\footnote{Here and in what follows, we adopt the reduced Planck mass unit, unless we specify it.}  
\begin{align}
    K = -\ln (-i(\tau -\Bar{\tau})) -\sum_i\ln (-i(T_i -\Bar{T}_i))
    -\sum_j\ln (-i(\tau_j -\Bar{\tau}_j)).
    \label{eq:Keff}
\end{align}
The superpotential of complex structure moduli and axio-dilaton 
can be generated by an existence of three-form fluxes~\cite{Gukov:1999ya},
\begin{align}
    W = (\tau -f \tau_3)g(\tau_1, \tau_2),
\end{align}
where we consider a particular form of the  three-form fluxes including $f$ and 
$g(\tau_1, \tau_2)$ is the proper function 
stabilizing $\tau_{1,2}$. 
The reason why we choose the above specific superpotential is that it leads to the massless direction in the $(\tau, \tau_3)$ moduli space 
at the supersymmetric Minkowski minimum~\cite{Hebecker:2017lxm,Kobayashi:2020hoc},
\begin{align}
    \partial_\tau W=\partial_{\tau_1} W=\partial_{\tau_2} W=\partial_{\tau_3} W=W=0,
\end{align}
where $g(\tau_1, \tau_2)$ is supposed to satisfy the above stabilization conditions,  for instance, $g(\tau_1, \tau_2)=(a^1\tau_1+a^2\tau_2)$ with $a^{1,2}$ being three-form flux quanta. 
As discussed later, the flat direction in the $(\tau, \tau_3)$ moduli space plays a crucial role of relating the QCD phase with the CKM phase. 
From the minimum $\tau=f \tau_3$, we define the flat direction ($\tau_f$) and 
the stabilized direction ($\tau_h$):
\begin{align}
    \tau_f &\equiv {\cal N}^{-1/2}\left(f\tau + \tau_3\right),
    \nonumber\\
    \tau_h &\equiv {\cal N}^{-1/2}\left(\tau - f\tau_3\right),    
\end{align}
with ${\cal N}=1+f^2$. 
Below the mass scale of stabilized $\tau_h$ with $\langle \tau_h \rangle =0$, 
the axio-dilaton and the complex structure moduli are described by the same 
modulus $\tau_f$ 
\begin{align}
    \tau &= {\cal N}^{-1/2}f\tau_f,
    \nonumber\\
    \tau_3 &= {\cal N}^{-1/2}\tau_f.
\label{eq:tautau3}
\end{align}

When magnetized D7-branes wrap the third torus $(T^2)_3$ and flavor structure of 
the quark sector is determined by the magnetic flux on $(T^2)_3$, the CKM phase 
is determined by $\tau_3$, namely $\tau_f$ below the mass scale of stabilized $\tau_h$. 
In this way, $\tau_f$ controls the magnitude of not only the QCD phase through Eq.~(\ref{eq:fD7}), 
but also the CKM phase through Eq.~(\ref{eq:Yukawa}). 
Note that K\"ahler axion ${\rm Re}(T)$ contributes to the QCD phase as in Eq.~(\ref{eq:fD7}), 
but in the following analysis, we assume that ${\rm Re}(T)$ is stabilized at the origin $\langle {\rm Re}(T)\rangle=0$ due to the non-perturbative effects for the K\"ahler moduli which 
enjoys a certain discrete symmetry for the axion. 
Such a simplification is useful to study the contribution from $\tau_f$ in the effective CP phase.  Furthermore, we  assume that the effective CP phase does not have the contributions from soft supersymmetry-breaking terms 
like gaugino masses and $B\mu$ terms to simplify our analysis. 
For that reason, we focus on the QCD and CKM phases, namely
\begin{align}
    \theta_{\rm eff} = \theta_{\rm QCD} 
    +{\rm Arg}\{{\rm Det}(y_uy_d)\},
\end{align}
and both are determined by the 
common axion $\tau_f$. In the following analysis, we explore the 
magnitudes of both CP phases on the basis of a concrete magnetized D-brane model. 

Finally, we comment on another possible scenario to entangle the QCD phase with the 
CKM phase. When there exist one-loop threshold corrections to the gauge kinetic 
function of D7-branes, the gauge kinetic function has a modular invariant function 
with respect to the complex structure moduli~\cite{Dixon:1990pc,Lust:2003ky,Blumenhagen:2006ci}. Then, both CP phases are 
related with each other. 
In this paper, we concentrate on the three-form flux scenario leading to a 
common origin of the QCD and CKM phases.

%%%%%%%%%%%%%%%%%%%%%%%%%%%%%%%%%%%%%%%%%%%%%%%%%%%%%%%%%%%%%%%%%%%%%%%%%%%%%%%%
\section{Concrete magnetized D-brane model}
\label{sec:3}
%%%%%%%%%%%%%%%%%%%%%%%%%%%%%%%%%%%%%%%%%%%%%%%%%%%%%%%%%%%%%%%%%%%%%%%%%%%%%%%%

To analyze the behavior of both the QCD and CKM phases in the moduli space 
of axion field $\tau_f$, we choose the specific magnetized D7-brane configuration wrapping the first and third torus and we assume that the flavor structure is only determined by the third torus $(T^2)_3$ on which the U(1) magnetic fluxes are inserted. In particular, we consider the toroidal orbifold $(T^2)_3/\mathbb{Z}_2$. 
Our purpose is to reveal whether there exists a moduli space of the axion field leading to the observed CKM phase consistent with the tiny strong CP phase or not. 
Therefore, we have not considered the global consistency conditions like tadpole 
cancellation conditions in this paper, since they depend on the existence of hidden sector 
as well as an amount of three-form fluxes. 

As discussed in Ref.~\cite{Abe:2008sx}, we start from U(8) super Yang-Mills action which can be regarded 
as the low-energy effective action of stacks of D7-branes. 
The magnetic fluxes are introduced to break U(8) to the standard-model gauge groups plus 
extra U(1)s. As displayed in Table~\ref{tab:1}, we assign the 
magnetic flux $n$ of quarks and Higgs fields and $\mathbb{Z}_2$ parity 
such that there exist three generations of quarks and five pairs of Higgs. 
Here, we choose the wrapping number of magnetized D7-branes $m^3=1$. 

\begin{table}[h]
\centering
    \begin{tabular}{|c|c|c|c|} \hline 
    \par  &  $Q_L$ & $Q_R$ & $H$ \\ \hline
	$n$ ~($\mathbb{Z}_2$ parity) &  $-5$\ ~(even) & $-7$\ ~(odd) & 12\ ~(odd) \\ \hline
%	Lepton & -4\ (even) & -8\ (odd) & 12\ (odd) \\ \hline
	\end{tabular}
	\caption{
		Magnetic fluxes for three generations of left-handed quarks $Q_L$, right-handed quarks $Q_R$ and five pairs of Higgs $H$.}
	\label{tab:1}
\end{table}

The Yukawa couplings in the quark sector 
\begin{align}
Y_{IJK}H_{K}Q_{LI}Q_{RJ}=
(Y_{IJ0}H_{0}+Y_{IJ1}H_{1}+Y_{IJ2}H_{2}+Y_{IJ3}H_{3}+Y_{IJ4}H_{4})Q_{LI}Q_{RJ},
\end{align}
are given by~\cite{Abe:2008sx}
\begin{align*}
Y_{IJ0}&=\frac{1}{\sqrt2}
\begin{pmatrix}
\eta_5-\eta_{65}&\eta_{185}-\eta_{115}&\sqrt2(\eta_{55}+\eta_{125})\\
\eta_{173}-\eta_{103}-\eta_{187}+\eta_{163}&\eta_{67}-\eta_{137}-\eta_{53}+\eta_{17}&
\eta_{113}-\eta_{43}-\eta_{127}+\eta_{197}\\
\eta_{79}-\eta_{149}-\eta_{19}+\eta_{89}&\eta_{101}-\eta_{31}-\eta_{199}+\eta_{151}&
\eta_{139}-\eta_{209}-\eta_{41}+\eta_{29}\\
\end{pmatrix},\\
Y_{IJ1}&=\frac1{\sqrt2}
\begin{pmatrix}
\eta_{170}-\eta_{110}&\eta_{10}-\eta_{130}&\sqrt2(\eta_{50}+\eta_{190})\\
\eta_{2}-\eta_{142}-\eta_{58}+\eta_{82}&\eta_{178}-\eta_{38}-\eta_{122}+\eta_{158}&
\eta_{62}-\eta_{202}-\eta_{118}+\eta_{22}\\
\eta_{166}-\eta_{26}-\eta_{194}+\eta_{94}&\eta_{74}-\eta_{206}-\eta_{46}+\eta_{94}&
\eta_{106}-\eta_{34}-\eta_{134}+\eta_{146}\\
\end{pmatrix},\\
Y_{IJ2}&=\frac1{\sqrt2}
\begin{pmatrix}
\eta_{75}-\eta_{135}&\eta_{165}-\eta_{45}&\eta_{15}-\eta_{195}\\
\eta_{173}-\eta_{33}-\eta_{117}+\eta_{93}&\eta_{3}-\eta_{207}-\eta_{123}+\eta_{87}&
\eta_{183}-\eta_{27}-\eta_{57}+\eta_{153}\\
\eta_{9}-\eta_{201}-\eta_{51}+\eta_{81}&\eta_{171}-\eta_{39}-\eta_{129}+\eta_{81}&
\eta_{69}-\eta_{141}-\eta_{111}+\eta_{99}\\
\end{pmatrix},\\
Y_{IJ3}&=\frac1{\sqrt2}
\begin{pmatrix}
\eta_{100}-\eta_{140}&\eta_{80}-\eta_{200}&\eta_{160}-\eta_{20}\\
\eta_{68}-\eta_{208}-\eta_{128}+\eta_{152}&\eta_{172}-\eta_{32}-\eta_{52}+\eta_{88}&
\eta_{8}-\eta_{148}-\eta_{188}+\eta_{92}\\
\eta_{184}-\eta_{44}-\eta_{124}+\eta_{164}&\eta_{4}-\eta_{136}-\eta_{116}+\eta_{164}&
\eta_{176}-\eta_{104}-\eta_{64}+\eta_{76}\\
\end{pmatrix},\\
Y_{IJ4}&=\frac1{\sqrt2}
\begin{pmatrix}
\eta_{145}-\eta_{205}&\eta_{95}-\eta_{25}&\eta_{85}-\eta_{155}\\
\eta_{107}-\eta_{37}-\eta_{47}+\eta_{23}&\eta_{73}-\eta_{143}-\eta_{193}+\eta_{157}&
\eta_{167}-\eta_{97}-\eta_{13}+\eta_{83}\\
\eta_{61}-\eta_{131}-\eta_{121}+\eta_{11}&\eta_{179}-\eta_{109}-\eta_{59}+\eta_{11}&
\eta_{1}-\eta_{71}-\eta_{181}+\eta_{169}\\
\end{pmatrix}
,
\end{align*}
up to the normalization factor.\footnote{In the following analysis, we focus on 
the mass ratios of quarks, elements of the CKM matrix and the Jarlskog invariant. 
It is then enough to omit the overall factors in Yukawa couplings, because the flavor structure is governed by the holomorphic Yukawa couplings.}
Now, we define
\begin{align}
    \eta_N \equiv \vartheta
\begin{bmatrix}
\frac{N}{M}\\
0
\end{bmatrix}
(0, {\cal N}^{-1/2}\tau_f M) 
\end{align}
with $M=420$ and $\tau_3={\cal N}^{-1/2}\tau_f$.

Before searching for the CKM phase compatible with the tiny strong CP phase, 
we discuss the axion-dependence of the Yukawa couplings in the next section.

%%%%%%%%%%%%%%%%%%%%%%%%%%%%%%%%%%%%%%%%%%%%%%%%%%%%%%%%%%%%%%%%%%%%%%%%%%%%%%%%
\subsection{CP phase from the Yukawa couplings}
%%%%%%%%%%%%%%%%%%%%%%%%%%%%%%%%%%%%%%%%%%%%%%%%%%%%%%%%%%%%%%%%%%%%%%%%%%%%%%%%

To reveal the functional behavior of the Yukawa couplings with respect to 
the axion $\tau_f$, we approximate the Jacobi-theta function with~\cite{Abe:2014vza}
\begin{align}
    \eta_N =\vartheta
\begin{bmatrix}
\frac{N}{M}\\
0
\end{bmatrix}
(0, {\cal N}^{-1/2}\tau_f M) 
\sim
e^{\frac{i\tau_f}{420}{\cal N}^{-1/2}N^2}
\end{align}
which is valid in the large complex structure limit 
${\rm Im}(\tau_3)={\cal N}^{-1/2}{\rm Im}(\tau_f)\gg 1$. 
Under this approximation, Yukawa couplings are expanded as
\begin{align}
\begin{split}
    Y_{IJ0} &= %\frac{a}{\sqrt{2}}
    \begin{pmatrix}
    \sqrt{2} \eta_{5} & -\sqrt{2} \eta_{115} & \sqrt{2} \eta_{55} \\
    -\eta_{103} & \eta_{17} & -\eta_{43} \\
    - \eta_{19} & -\eta_{31} & \eta_{29}
    \end{pmatrix}
    ,\\
    Y_{IJ1} &= %\frac{a}{\sqrt{2}}
    \begin{pmatrix}
    -\sqrt{2} \eta_{110} & \sqrt{2} \eta_{10} & \sqrt{2} \eta_{50} \\
    \eta_{2} & -\eta_{38} & \eta_{22} \\
    - \eta_{26} & -\eta_{46} & - \eta_{34}
    \end{pmatrix}
    ,\\
    Y_{IJ2} &= %\frac{a}{\sqrt{2}}
    \begin{pmatrix}
    \sqrt{2} \eta_{75} & -\sqrt{2} \eta_{45} & - \sqrt{2} \eta_{15} \\
    -\eta_{33} & -\eta_{3} & -\eta_{27} \\
     \eta_{9} & -\eta_{39} & - \eta_{69}
    \end{pmatrix}
    ,\\
    Y_{IJ3} &= %\frac{a}{\sqrt{2}}
    \begin{pmatrix}
    \sqrt{2} \eta_{100} & \sqrt{2} \eta_{80} & - \sqrt{2} \eta_{20} \\
    \eta_{68} & - \eta_{32} & \eta_{8} \\
    - \eta_{44} & \eta_{4} & - \eta_{64}
    \end{pmatrix}
    ,\\
Y_{IJ4} &= %\frac{a}{\sqrt{2}}
    \begin{pmatrix}
    \sqrt{2} \eta_{145} & - \sqrt{2} \eta_{25} & \sqrt{2} \eta_{85} \\
    \eta_{23} & \eta_{73} & - \eta_{13} \\
    \eta_{11} & \eta_{11} & \eta_{1}
    \end{pmatrix}
.
    \end{split}
    \label{eq:YK}    
\end{align}

Recalling that ${\rm Arg}\{{\rm Det}(y_uy_d)\}$ has the following property
\begin{align}
   {\rm Arg}\{{\rm Det}(y_uy_d)\} = {\rm Arg}\{{\rm Det}(y_u)\} +{\rm Arg}\{{\rm Det}(y_d)\},
\end{align}
for non-zero complex numbers ${\rm Det}(y_u),{\rm Det}(y_d)$, 
the functional behavior of the CKM phase with respect to 
the axion can be understood by evaluating the ${\rm Det}(Y_K)$ with $K=0,1,2,3,4$. In the large complex structure limit ${\rm Im}(\tau_3)={\cal N}^{-1/2}{\rm Im}(\tau_f) \gg 1$, the approximate form of ${\rm Det}(Y_K)$ is given by
\begin{align}
    {\rm Det}(Y_{IJ0}) &\sim e^{11i\pi \tau_3/4}\left(1-e^{4i\pi\tau_3}\right),\qquad
    {\rm Det}(Y_{IJ1}) \sim -e^{11i\pi \tau_3}\left(1 +e^{24i\pi\tau_3}\right),\nonumber\\
    {\rm Det}(Y_{IJ2}) &\sim e^{3i\pi\tau_3/4}\left(-1+2e^{6i\pi\tau_3}\right),\qquad
    {\rm Det}(Y_{IJ3}) \sim e^{8i\pi\tau_3}\left(1-e^{4i\pi\tau_3}\right),
    \nonumber\\
    {\rm Det}(Y_{IJ4}) &\sim e^{61i\pi\tau_3/28}\left(1+e^{4i\pi\tau_3/7}\right),    
\end{align}
from which ${\rm Arg}\{{\rm Det}(Y_{IJK})\}$ is a non-linear function of the axion ${\rm Re}(\tau_3)$, rather than the linear function. 

\begin{figure}[htbp]
\centering
  \begin{minipage}{0.41\columnwidth}
  \centering
     \includegraphics[scale=0.35]{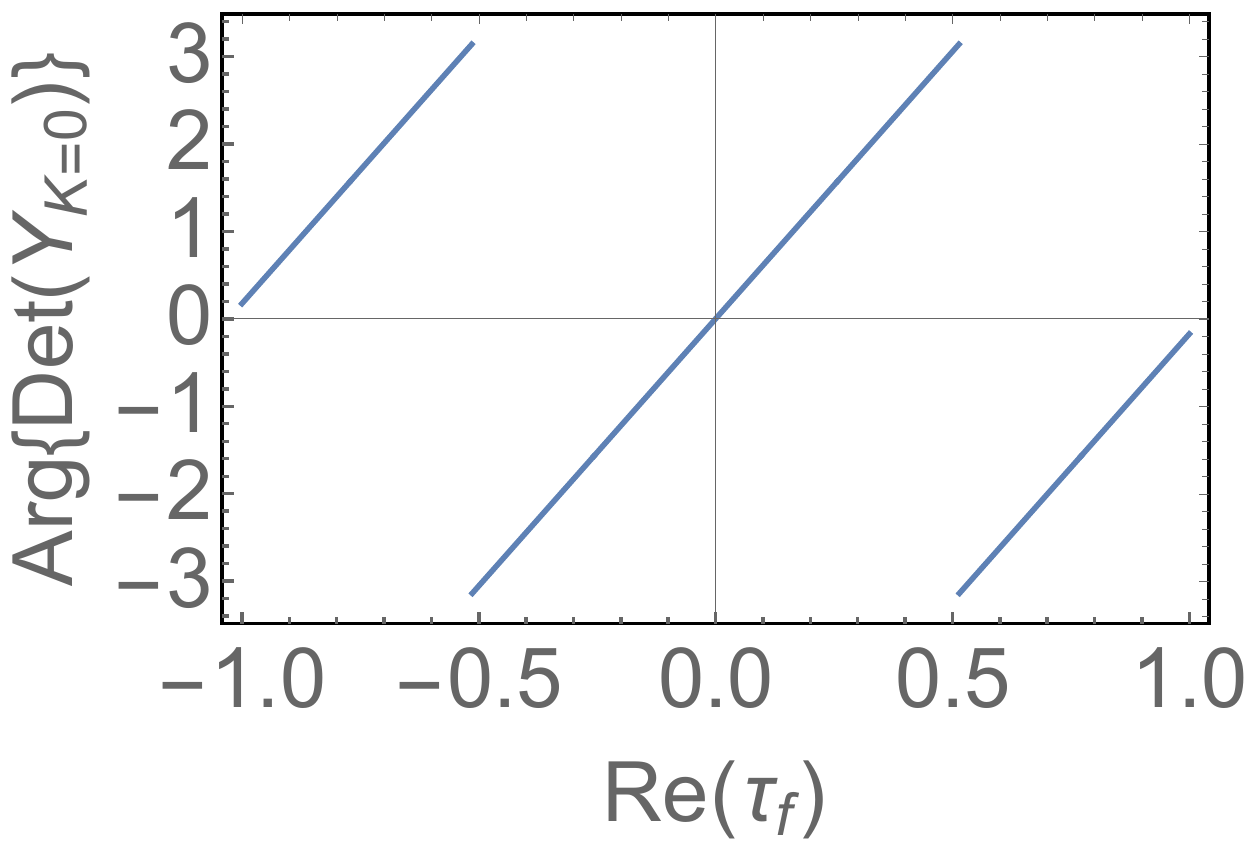}
  \end{minipage}
  \begin{minipage}{0.4\columnwidth}
  \centering
     \includegraphics[scale=0.35]{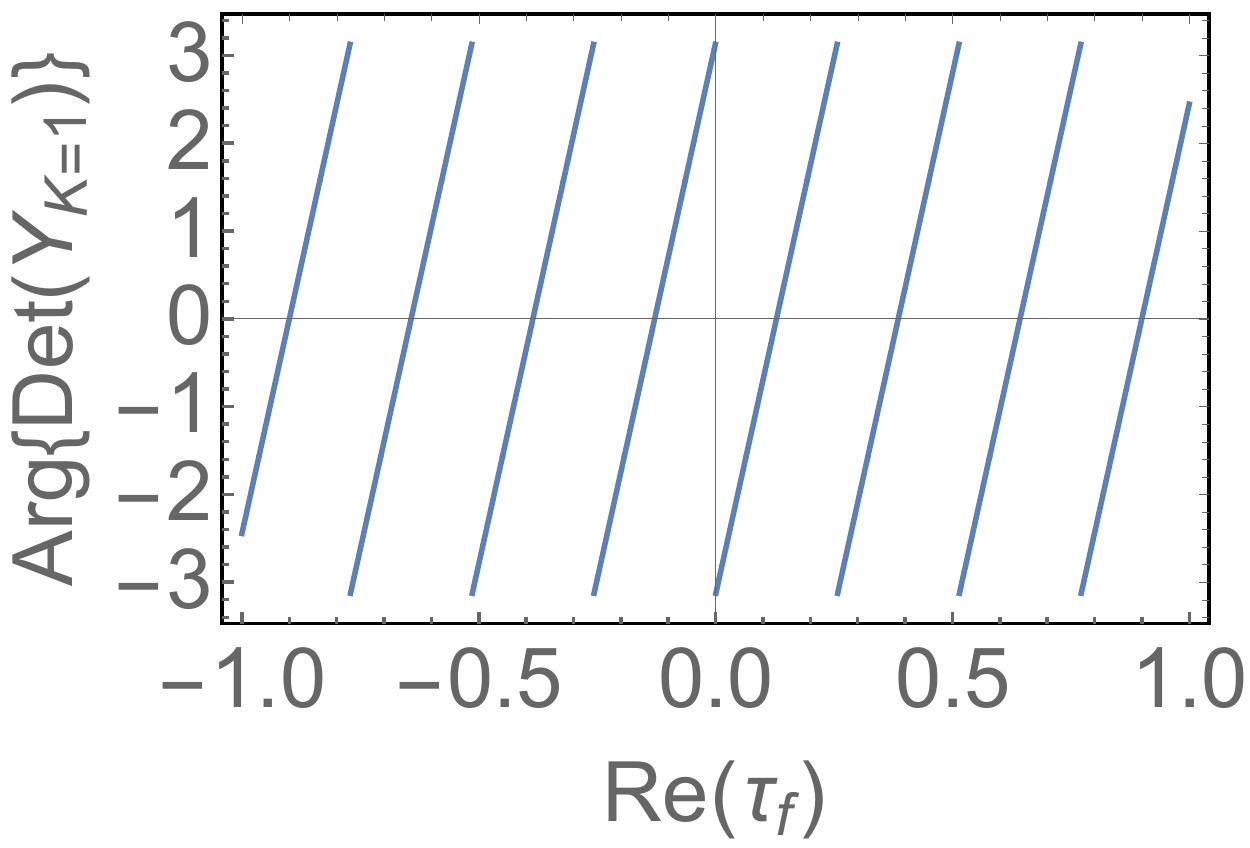}
  \end{minipage} \\
  \begin{minipage}{0.41\columnwidth}
  \centering
     \includegraphics[scale=0.35]{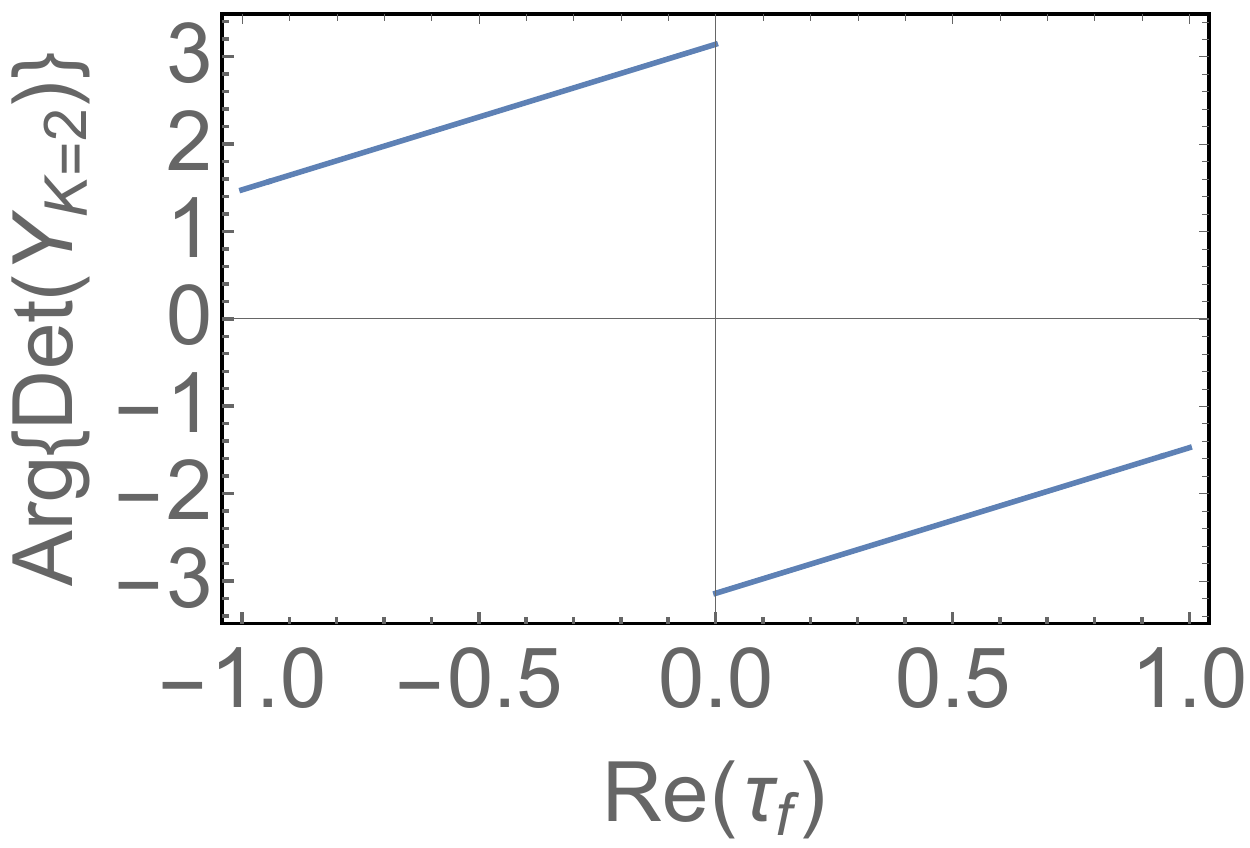}
  \end{minipage}
  \begin{minipage}{0.4\columnwidth}
  \centering
     \includegraphics[scale=0.35]{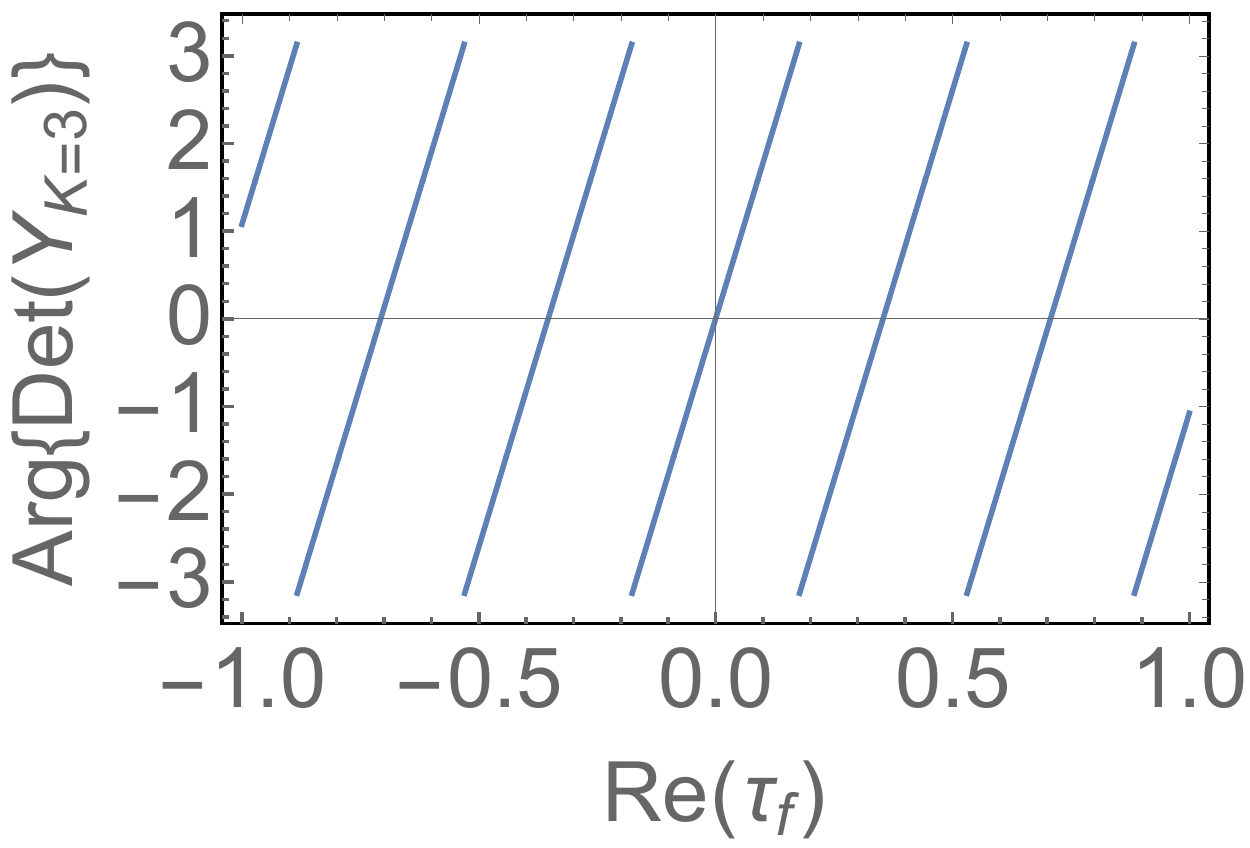}
  \end{minipage}
  \begin{minipage}{0.4\columnwidth}
  \centering
     \includegraphics[scale=0.35]{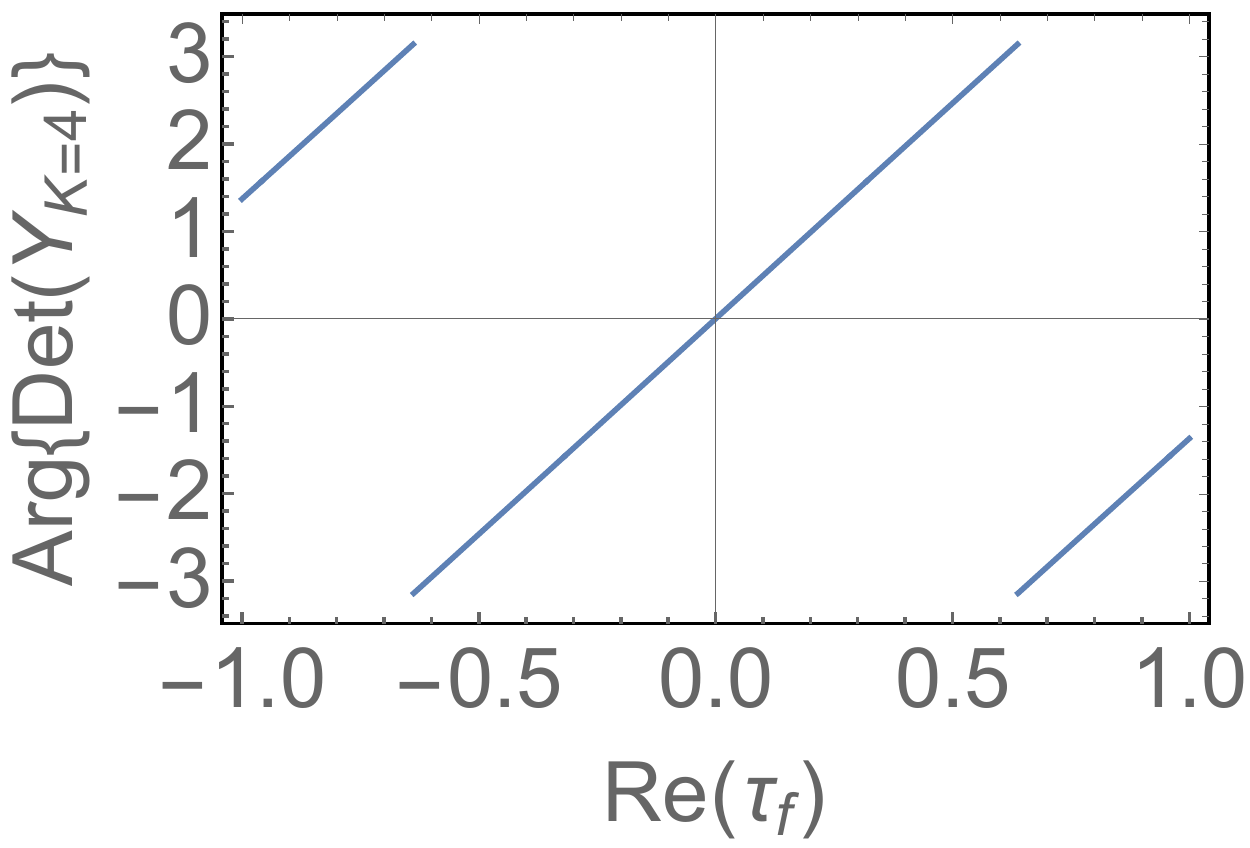}
  \end{minipage}
\caption{Plots of ${\rm Arg}\{{\rm Det}(Y_{IJK})\}$ with $K=0,1,2,3,4$ by setting $f=1$ and ${\rm Im}(\tau_f)=2$.}
    \label{fig:YKapp}
\end{figure}

Indeed, Figure~\ref{fig:YKapp} shows that ${\rm Arg}\{{\rm Det}(Y_{IJK})\}$ employing Eq.~(\ref{eq:YK}) is a complicated 
function of ${\rm Re}(\tau_f)$, where we set ${\rm Im}(\tau_f)=2$ and $f=1$. The functional behavior is the same even when we set the 
other ${\rm Im}(\tau_f)$ and $f$. Such an axion dependence is a consequence 
of the definition of the argument Arg$\{ {\rm Det}(z)\}$ with $z$ being a complex number:
\begin{align}
    {\rm Arg}\{ {\rm Det}(z)\} =\left\{
\begin{array}{c}
        {\rm Arctan}\left( \cfrac{{\rm Im}({\rm Det}(z))}{{\rm Re}({\rm Det}(z))}\right)\qquad ({\rm Re}({\rm Det}(z))>0,{\rm Im}({\rm Det}(z))\gtrless 0)\\
        {\rm Arctan}\left( \cfrac{{\rm Im}({\rm Det}(z))}{{\rm Re}({\rm Det}(z))}\right)+\pi\qquad ({\rm Re}({\rm Det}(z))<0,{\rm Im}({\rm Det}(z))>0)\\
        {\rm Arctan}\left( \cfrac{{\rm Im}({\rm Det}(z))}{{\rm Re}({\rm Det}(z))}\right)-\pi\qquad ({\rm Re}({\rm Det}(z))<0,{\rm Im}({\rm Det}(z))<0)\\
\end{array}
\right.
.
\label{eq:Arg}
\end{align}

%%%%%%%%%%%%%%%%%%%%%%%%%%%%%%%%%%%%%%%%%%%%%%%%%%%%%%%%%%%%%%%%%%%%%%%%%%%%%%%%
%%%%%%%%%%%%%%%%%%%%%%%%%%%%%%%%%%%%%%%%%%%%%%%%%%%%%%%%%%%%%%%%%%%%%%%%%%%%%%%%

%%%%%%%%%%%%%%%%%%%%%%%%%%%%%%%%%%%%%%%%%%%%%%%%%%%%%%%%%%%%%%%%%%%%%%%%%%%%%%%%
\subsection{Suppressed CP phase}
%%%%%%%%%%%%%%%%%%%%%%%%%%%%%%%%%%%%%%%%%%%%%%%%%%%%%%%%%%%%%%%%%%%%%%%%%%%%%%%%

In this section, we take into account the QCD phase $\theta_{\rm QCD}$ in addition 
to the CP phase from Yukawa couplings treated in the previous section and check whether there exists 
the axionic moduli space to realize the tiny strong CP phase and the ${\cal O}(1)$ 
CKM phase. 
The QCD phase is now determined by
\begin{align}
    \theta_{\rm QCD} = M_{\rm QCD} \tau = M_{\rm QCD}{\cal N}^{-1/2}f\tau_f,
\end{align}
originating from Eq.~(\ref{eq:fD7}) with Eq.~(\ref{eq:tautau3}) by 
assuming ${\rm Re}(T)=0$ for the K\"ahler axion. Here, we denote the 
magnetic flux contributions by $M_{\rm QCD}$. 
To evaluate the magnitude of the CKM phase, we examine the 
Jarlskog invariant ($J$)
\begin{align}
    J\sum_{m,n=1}^3\epsilon_{ikm}\epsilon_{jln}= {\rm Im}\left[ V_{ij}V_{kl}V_{il}^\ast V_{kj}^\ast\right],
\end{align}
where $V_{ij}$ is the element of the CKM matrix and $\epsilon_{ikm}$ is 
the Levi-Civita symbol. 

For an illustrative purpose, we adopt the specific mass matrices for 
up- and down-type quarks:
\begin{align}
    M_u &= Y_{IJ4}\langle H_{u,4}\rangle +Y_{IJ3}\langle H_{u,3}\rangle 
         = \langle H_{u,4}\rangle \left(Y_{IJ4} +Y_{IJ3}\rho_u \right),
    \nonumber\\
    M_d &= Y_{IJ4}\langle H_{d,4}\rangle +Y_{IJ3}\langle H_{d,3}\rangle 
        =\langle H_{d,4}\rangle\left(Y_{IJ4} +Y_{IJ3}\rho_d \right),
\label{eq:MuMd}
\end{align}
meaning that up- and down-type Higgs $H_{u,d}$ are linear combinations of $H_4$ and 
$H_3$. Here and in what follows, we assume that both Higgs fields are assumed to be nonvanishing real field values and for our purpose, the vacuum expectation values of Higgs fields are parametrized by
\begin{align}
    \rho_u = \frac{\langle H_{u,3}\rangle }{\langle H_{u,4}\rangle},\qquad
    \rho_d = \frac{\langle H_{d,3}\rangle }{\langle H_{d,4}\rangle}.
\end{align}
The overall factors $\langle H_{u(d),4}\rangle$ in Eq.~(\ref{eq:MuMd}) are assumed to 
realize the scale of quark masses. 
The reason why we adopt $Y_{IJ4}$ and $Y_{IJ3}$ for the quark mass matrices is that they have 
a hierarchical structure among three generations of quarks as analytically discussed in Ref.~\cite{Abe:2014vza}. 
The above form of quark mass matrices leads to the following CP 
phase in the large complex structure limit $\tau_3\gg 1$,
\begin{align}
    {\rm Arg}\{{\rm Det} (y_uy_d)\}
    \simeq \sum_{i=u,d} \sqrt{2}e^{27i\pi \tau_3/14}\left(e^{i\pi\tau_3/4}+e^{23i\pi\tau_3/28}-\rho_i -e^{4i\pi\tau_3/7}\rho_i -e^{9i\pi\tau_3/28}\rho_i^2 \right),
\end{align}
where $\rho_{u,d}$ is assumed to be ${\cal O}(1)$. 

\begin{figure}[h]
\centering
  \begin{minipage}{0.4\columnwidth}
  \centering
     \includegraphics[scale=0.45]{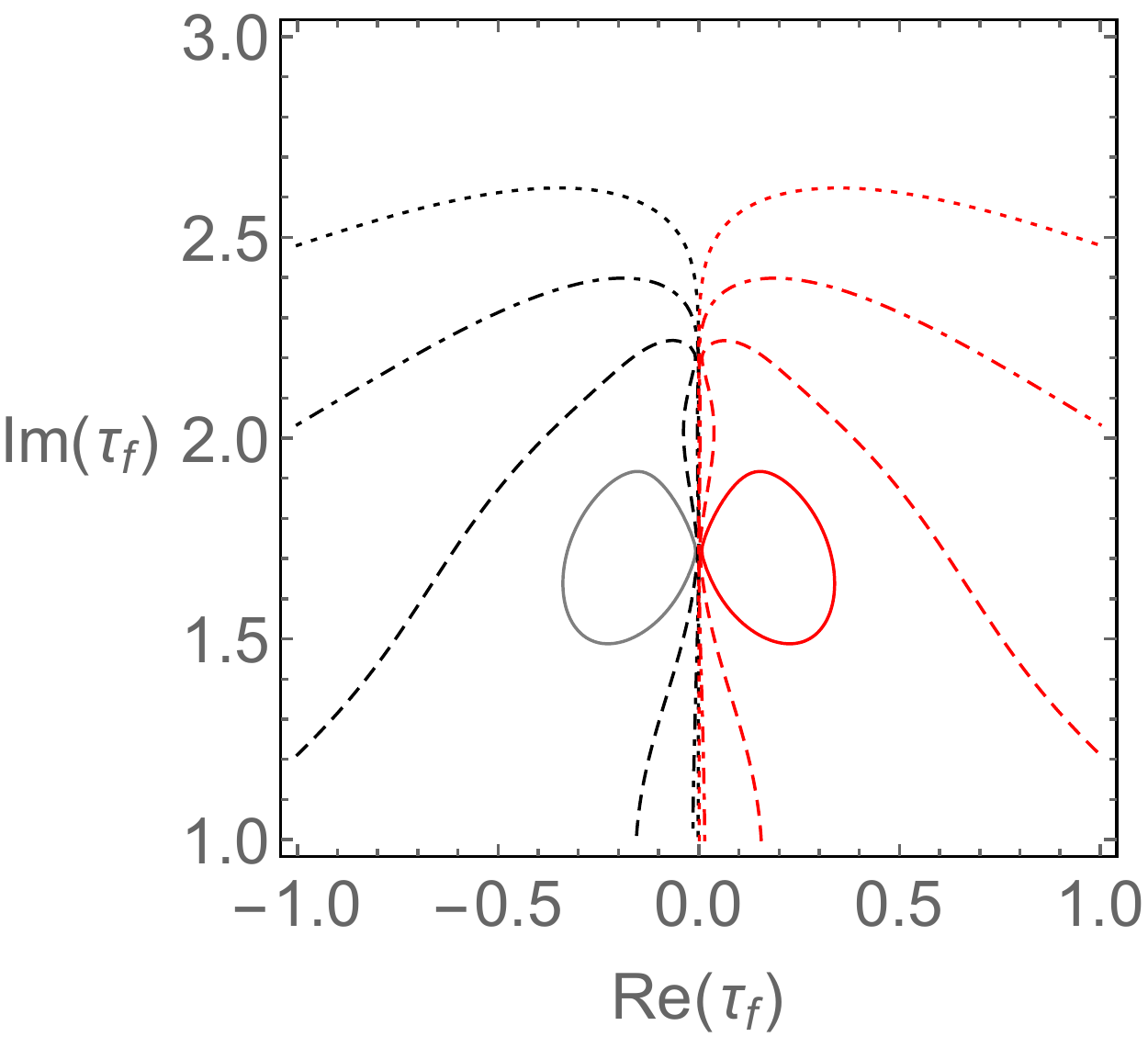}
\end{minipage}
  \begin{minipage}{0.4\columnwidth}
  \centering
     \includegraphics[scale=0.45]{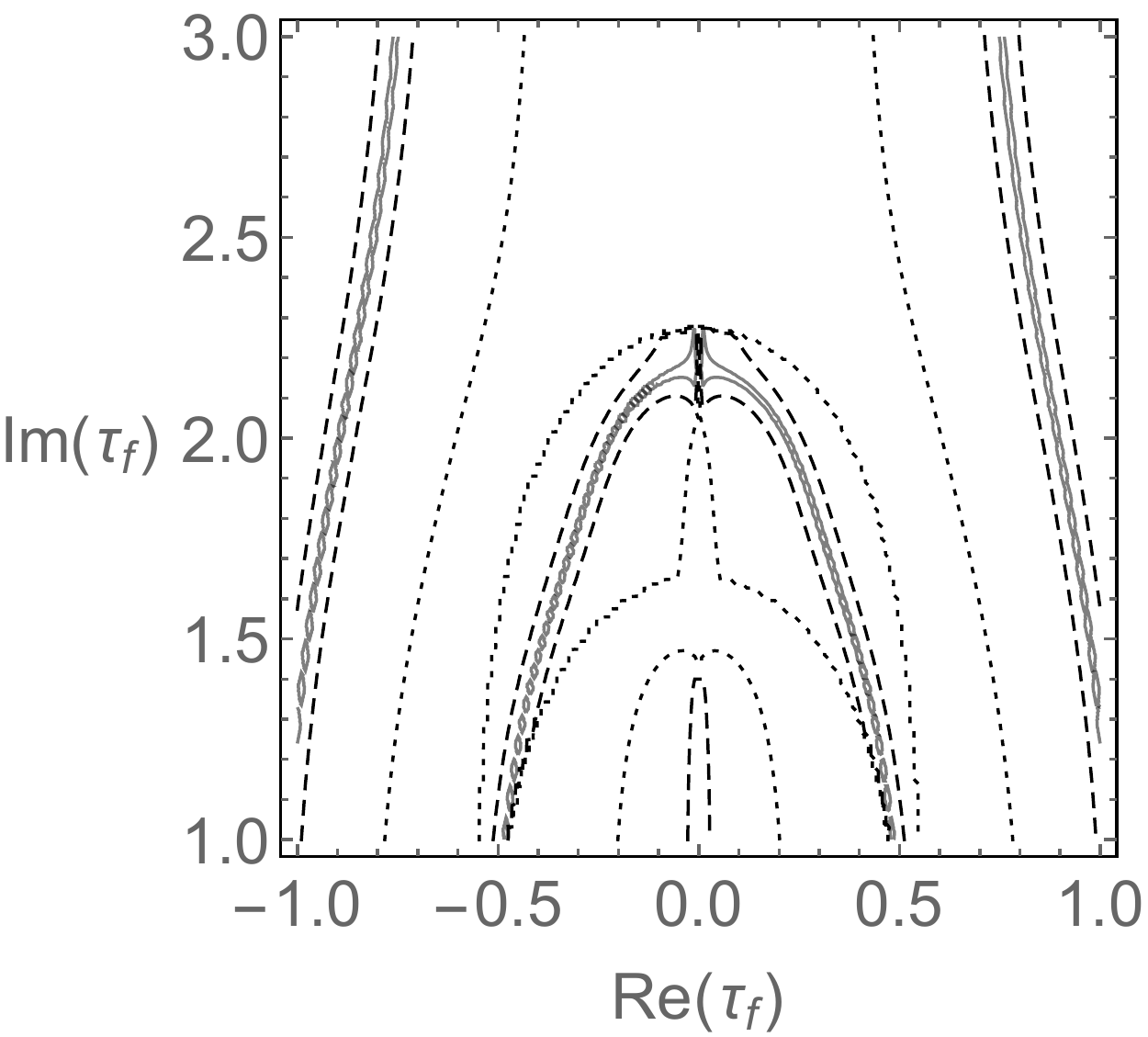}
\end{minipage}
\caption{The functional behavior of the Jarlskog invariant $J$ in the left panel and the effective CP phase Log($|\theta_{\rm eff}|$) in the right panel with respect to ${\rm Re}(\tau_f)$ and ${\rm Im}(\tau_f)$, where the parameters are set as in Eq.~(\ref{eq:parameters}). In the left panel, black (red) dotted, dotdashed, dashed and solid curves correspond to $J=10^{-7},10^{-6},10^{-5},3\times 10^{-5}$ ($J=-10^{-7},-10^{-6},-10^{-5},-3\times 10^{-5}$), respectively.  
In the right panel, black dotted, dashed and solid curves correspond to ${\rm Log}(|\theta_{\rm eff}|)=1, -1, -3$, respectively.}
    \label{fig:CPeff}
\end{figure}

\begin{figure}[htbp]
\centering
     \includegraphics[scale=0.3]{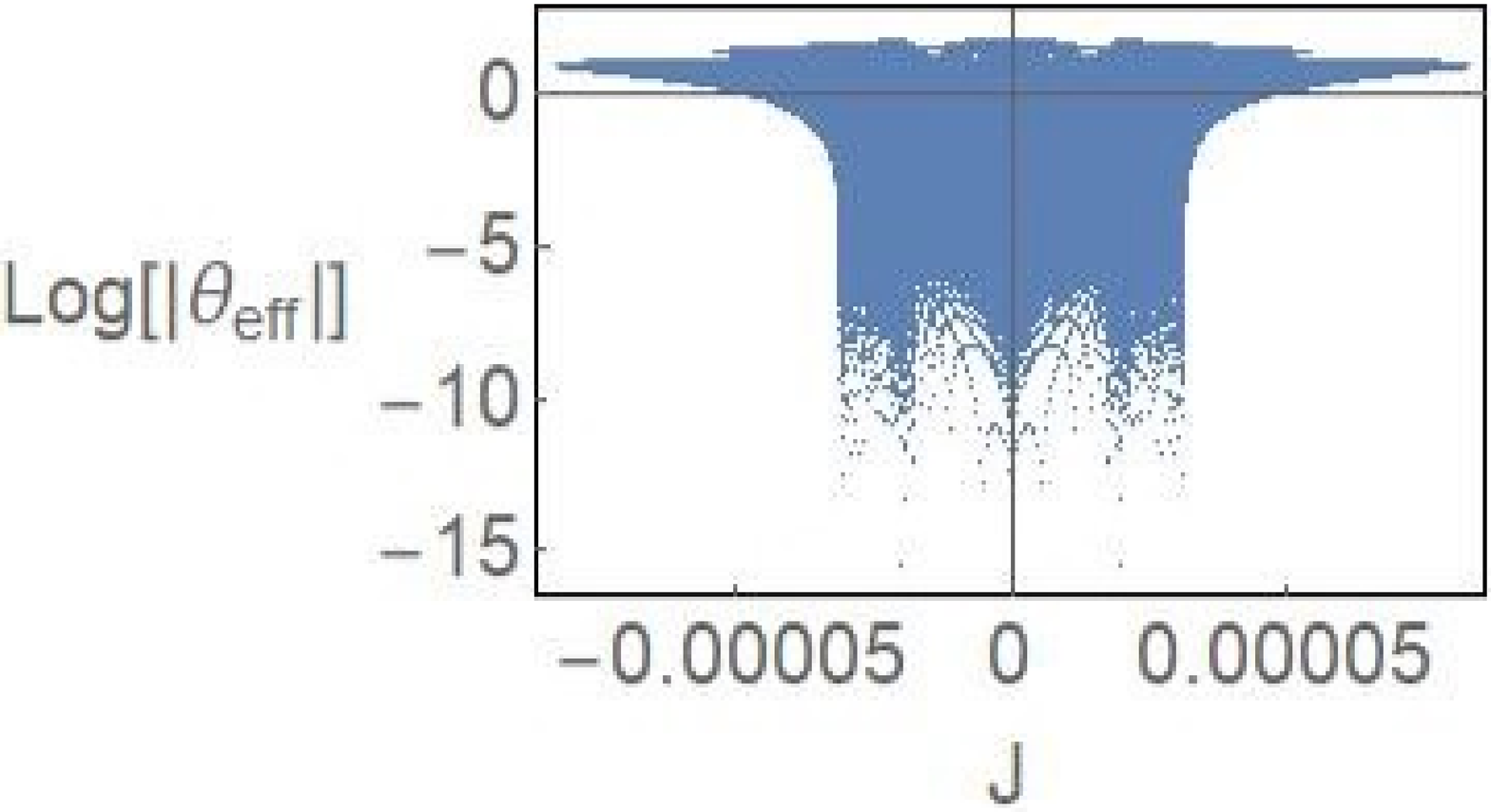}
\caption{The functional behavior of the Jarlskog invariant $J$ versus the effective CP phase Log($|\theta_{\rm eff}|$) within $-1/2 \leq{\rm Re}(\tau_f)\leq 1/2$ and $1 \leq {\rm Im}(\tau_f)\leq 2.5$ with the step size $5\times 10^{-4}$, where the parameters 
are the same with Figure~\ref{fig:CPeff}. 
When the step size is narrower and narrower, the 
effective CP phase $\theta_{\rm eff}$ is close to 0 at the specific 
value of the Jarlskog invariant.}
    \label{fig:CPvsJ}
\end{figure}

Following the above setup, we numerically estimate the Jarlskog invariant $J$ and the effective CP phase $\theta_{\rm eff}$ as functions of ${\rm Re}(\tau_f)$ and ${\rm Im}(\tau_f)$ in Figure~\ref{fig:CPeff}, where we choose the following parameters
\begin{align}
    f=1, \qquad M_{\rm QCD}=1,\qquad
    \rho_u =0.3,\qquad \rho_d =0.4.
\label{eq:parameters}
\end{align}
From Figure~\ref{fig:CPeff}, the effective CP phase vanishes periodically in the axionic direction due to the property of 
the arctangent function in Eq.~(\ref{eq:Arg}) and at the same time, a small but finite Jarlskog invariant 
$J$ can be realized at the minimum with $\theta_{\rm eff}= 0$ as shown in 
Figure~\ref{fig:CPvsJ}. 
Indeed, semi-realistic values of the mass ratios for quarks, elements of the CKM matrix and the Jarlskog 
invariant $J$ are obtained at the benchmark point in Table~\ref{tb:model}. 
This result is a consequence of the non-trivial axion-dependent function of ${\rm Arg}\{{\rm Det}(y_uy_d)\}$. 
For instance, when the Yukawa couplings have a FN-type, $\theta_{\rm eff}$ is a linear function of the 
axion, indicating that vanishing $\theta_{\rm eff}$ is occurred at ${\rm Re}(\tau_f)=0$. 
Since the nonvanishing CKM phase is induced by a nonzero value of ${\rm Re}(\tau_f)$, we cannot 
obtain a nonzero Jarlskog invariant in a FN-type scenario. 
As a result, the important point to realize a small but finite $J$ at $\theta_{\rm eff}=0$ is the 
non-trivial axion dependent function of the Yukawa couplings. 
In this paper, we assume a proper mechanism to realize $\theta_{\rm eff}=0$ by non-perturbative effects 
in a hidden sector at a scale larger than the electroweak scale. 
If the hidden sector also involves the axion-dependent CP phase from the Yukawa couplings in addition to the CP phase 
from the gauge kinetic function, it would lead to the observed value of the Jarlskog invariant 
at $\theta_{\rm eff}=0$. We hope to report on this in a future work. 
Furthermore, we focus on the Yukawa couplings of quarks living on magnetized D-branes wrapping tori for technical reason, but it is interesting to explore more general background like Calabi-Yau orientifolds.

%%%%%%%%%%%%%%%%%%%%%%%%%%%%%%%%%%%%%%%%%%%%%%%%%%%%%%%%%%%%%%%%%%%%%%%%%%%%%%%%
%%%%%%%%%%%%%%%%%%%%%%%%%%%%%%%%%%%%%%%%%%%%%%%%%%%%%%%%%%%%%%%%%%%%%%%%%%%%%%%%
%%%%%%%%%%%%%%%%%%%%%%%%%%%%%%%%%%%%%%%%%%%%%%%%%%%%%%%%%%%%%%%%%%%%%%%%%%%%%%%%
%%%%%%%%%%%%%%%%%%%%%%%%%%%%%%%%%%%%%%%%%%%%%%%%%%%%%%%%%%%%%%%%%%%%%%%%%%%%%%%%

%%%%%%%%%%%%%%%%%%%%%%%%%%%%%%%%%%%%%%%%%%%%%%%%%%%%%%%%%%%%%%%%%%%%%%%%%%%%%%%%
\begin{table}[htbp]
\centering
%\scalebox{0.9}{
    \begin{tabular}{|c|c|c|} \hline 
    \par & Benchmark values & Observed values \\ \hline
	$(m_u, m_c, m_t)/m_t$ & $(5.7 \times 10^{-4}, 1.2 \times 10^{-2}, 1)$ & $(6.5 \times 10^{-6}, 3.2 \times 10^{-3}, 1)$ \\ \hline
	$(m_d, m_s, m_b)/m_b$ & $(9.8 \times 10^{-4}, 2.0 \times 10^{-2}, 1)$ & $(1.1 \times 10^{-3}, 2.2 \times 10^{-2}, 1)$ \\ \hline
	$|V_{CKM}|$
	&$\begin{pmatrix}
	0.98 & 0.19 & 0.0054 \\
	0.19 & 0.98 & 0.035 \\
	0.0038 & 0.036 & 1.0
	\end{pmatrix}$
	&$\begin{pmatrix}
	0.97 & 0.22 & 0.0037 \\
	0.22 & 0.97 & 0.042 \\
	0.0090 & 0.041 & 1.0
	\end{pmatrix}$
	\\ \hline
	$J$
	& 
	$1.98\times 10^{-5}$
	&
	$3.18\times 10^{-5}$
	\\ \hline
	\end{tabular}
	\caption{
		The mass ratios for quarks, elements of the CKM matrix and the Jarlskog invariant $J$ at the benchmark point, 
		where we set ${\rm Re}(\tau_f)\simeq -0.2188$ and ${\rm Im}(\tau_f)=2$ leading to $\theta_{\rm eff}\simeq 0$ and parameters are chosen as in Eq.~(\ref{eq:parameters}). 
		Here, we use the GUT scale running masses for the observed values~\cite{Xing:2007fb} and 
		the value of the Jarlskog invariant in Ref.~\cite{Tanabashi:2018oca}.}
	\label{tb:model}
\end{table}

%%%%%%%%%%%%%%%%%%%%%%%%%%%%%%%%%%%%%%%%%%%%%%%%%%%%%%%%%%%%%%%%%%%%%%%%%%%%%%%%

\clearpage
%%%%%%%%%%%%%%%%%%%%%%%%%%%%%%%%%%%%%%%%%%%%%%%%%%%%%%%%%%%%
%%%%%%%%%%%%%%%%%%%%%%%%%%%%%%%%%%%%%%%%%%%%%%%%%%%%%%%%%%%%
\section{Conclusions}
%%%%%%%%%%%%%%%%%%%%%%%%%%%%%%%%%%%%%%%%%%%%%%%%%%%%%%%%%%%%
%%%%%%%%%%%%%%%%%%%%%%%%%%%%%%%%%%%%%%%%%%%%%%%%%%%%%%%%%%%%

From the view point of string theory, the strong CP and CKM 
phases are not constants, but they are determined by the axion fields 
originated from the higher-dimensional gauge fields. 
It is then natural to ask whether both phases have a common 
origin and at the same time, the observed value of the CKM 
phase is compatible with almost vanishing strong CP phase or not. 

In this paper, we first proposed the mechanism to relate the 
strong CP phase with the CKM phase in Type IIB flux vacua 
with magnetized D7-branes. 
We demonstrated that the axio-dilaton appearing in the gauge kinetic 
function and the complex structure moduli in Yukawa couplings on magnetized 
D7-branes are entangled by certain three-form fluxes which lead 
to the massless direction in the moduli spaces of the axion-dilaton 
and one of the complex structure moduli. 
Note that it is possible to have a common axion field 
associated with the complex structure moduli in 
the gauge kinetic function and Yukawa couplings through 
one-loop threshold corrections to the gauge kinetic function~\cite{Dixon:1990pc,Lust:2003ky,Blumenhagen:2006ci}. 

To estimate the value of the CKM phase, we examine the Yukawa couplings 
on magnetized D-branes wrapping tori on which analytical calculation has 
been performed in Ref.~\cite{Cremades:2004wa}. 
It is known that the CP phase is induced by the nonvanishing axion field. 
If the CP phase ${\rm Arg}\{{\rm Det}(y_uy_d)\}$ is linearly dependent of 
the axion as in the Froggatt-Nielsen model~\cite{Froggatt:1978nt}, the strong CP phase 
$\theta_{\rm eff}$ becomes zero at the origin of the axion field. 
However, thanks to the non-trivial axion-dependent function of the CP phase ${\rm Arg}\{{\rm Det}(y_uy_d)\}$ on toroidal background with magnetic fluxes, we find that 
observed value of the Jarlskog invariant is consistent with the vanishing 
strong CP phase. 
In this paper, we focus on the bare CP phases, but radiative corrections as well as the 
supersymmetry-breaking effects give rise to nonvanishing CP phases in general, 
which will be one of the important future work. 
%Furthermore, we assume the stabilization of axion field by certain non-perturbative 
%dynamics, which can be determined once the amount of magnetic and 
%three-form fluxes are fixed. 
Furthermore, we assume the stabilization of axion field by certain non-perturbative dynamics in a hidden sector. 
We will leave the detailed axion stabilization for a future work. 
The relation between the strong CP and CKM phases would be possible for not only 
the toroidal orientifold background in Type IIB string context, but also more general Calabi-Yau orientifolds in other superstring theory, for instance, Type IIA intersecting D6-brane system and heterotic line bundle models. 
This is because three-form fluxes give rise to the massless direction in 
the moduli spaces of the axio-dilaton and the complex structure moduli. Both play 
an important role of determining the gauge kinetic function as well as the Yukawa 
couplings. 
We will report on this interesting work in the future.

%%%%%%%%%%%%%%%%%%%%%%%%%%%%%%%%%%%%%%%%%%%%%%%%%%%%%%%%%%%%
%%%%%%%%%%%%%%%%%%%%%%%%%%%%%%%%%%%%%%%%%%%%%%%%%%%%%%%%%%%%
\subsection*{Acknowledgements}
%%%%%%%%%%%%%%%%%%%%%%%%%%%%%%%%%%%%%%%%%%%%%%%%%%%%%%%%%%%%
%%%%%%%%%%%%%%%%%%%%%%%%%%%%%%%%%%%%%%%%%%%%%%%%%%%%%%%%%%%%

%We would like to thank XXX for useful discussions and comments. 
T. K. was supported in part by MEXT KAKENHI Grant Number JP19H04605. 
H. O. was supported by Grant-in-Aid for JSPS Research Fellows No. 19J00664.

\end{document}